\documentclass{ws-p8-50x6-00}

\begin{document}
\tolerance=10000
\hbadness=10000

\title{Mean-field nuclear structure calculations
on a Basis-Spline Galerkin lattice}

\author{Volker E. Oberacker and A. Sait Umar}

\address{Dept. Physics \& Astronomy, Vanderbilt University,
Nashville, TN 37235, USA\\E-mail: volker.e.oberacker@vanderbilt.edu}


\maketitle

\abstracts{
Our goal is to carry out high-precision nuclear structure calculations
in connection with Radioactive Ion Beam Facilities. The main challenge
for the theory of drip line nuclei is that the outermost nucleons are
weakly bound (implying a large spatial distribution) and that these
states are strongly coupled to the particle continuum. For these
reasons, the traditional basis expansion methods fail to converge. We
overcome these problems by representing the nuclear Hamiltonian on a
lattice utilizing the Galerkin method with Basis-Spline test
functions. We discuss tests of the numerical method and applications
to the deformed shell model, HF+BCS and HFB mean field theories.
}


\section{Introduction}
In recent years, the area of nuclear structure physics has shown
substantial progress and rapid growth \cite{NSAC96,ISOL97}. With
detectors such as GAMMASPHERE and EUROGAM, the limits of total angular
momentum and deformation in atomic nuclei have been explored, and new
neutron rich nuclei have been identified in spontaneous fission
studies. Gamma-ray detectors under development such as GRETA
\cite{IYL98} will have improved resolving power and should allow for
the identification of weakly populated states never seen before in
nuclei. Particularly exciting is the proposed construction of a
next-generation ISOL FACILITY in the United States which has been been
identified in the 1996 DOE/NSAC Long Range Plan \cite{NSAC96} as the
highest priority for major new construction.

These experimental developments as well as recent advances in
computational physics have sparked renewed interest in nuclear
structure theory. In contrast to the well-understood behavior near the
valley of stability, there are many open questions as we move towards
the proton and neutron driplines and towards the limits in mass number
(superheavy region). While the proton dripline has been explored
experimentally up to Z=83, the neutron dripline represents mostly
``terra incognita''. In these exotic regions of the nuclear chart, one
expects to see several new phenomena: near the neutron dripline, the
neutron-matter distribution will be very diffuse and of large size
giving rise to ``neutron halos'' and ``neutrons skins''. One also
expects new collective modes associated with this neutron skin, e.g.
the ``scissors'' vibrational mode or the ``pygmy'' resonance. In
proton-rich nuclei, we have recently seen both spherical and deformed
proton emitters; this ``proton radioactivity'' is  caused by the
tunneling of weakly bound protons through the Coulomb barrier.
The investigation of the properties of exotic nuclei is also essential
for our understanding of nucleosynthesis in stars and stellar
explosions (rp- and r-process). Our primary goal is to carry out
high-precision nuclear structure calculations in connection with
Radioactive Ion Beam Facilities. Some of the topics of interest are
the effective N-N interaction at large isospin, large pairing
correlations and their density dependence, neutron halos/skins, and
proton radioactivity. Specifically, we are interested in calculating
observables such as the total binding energy, charge radii, densities
$\rho_{p,n}({\bf r})$, separation energies for neutrons and protons,
pairing gaps, and potential energy surfaces.

There are many theoretical approaches to nuclear structure physics.
For lack of space, we mention only three of these: in the macroscopic
- microscopic method, one combines the liquid drop / droplet model
with a microscopic shell correction from a deformed single-particle
shell model (M\"oller and Nix \cite{MN97}, Nazarewicz et al.
\cite{NW94}). For relatively light nuclei, it is possible to
diagonalize the nuclear Hamiltonian in a shell model basis. Barrett et
al. \cite{NB97} have recently carried out large-basis no-core shell
model calculations for p-shell nuclei. A different approach to the
interacting nuclear shell model is the Shell Model Monte Carlo (SMMC)
method developed by Dean et al. \cite{KDL97} which does not involve
matrix diagonalization but a path integral over auxiliary fields. This
method has been applied to fp-shell and medium-heavy nuclei.
Finally, for heavier nuclei one utilizes either nonrelativistic
\cite{DF84,DN96,RB97} or relativistic \cite{ND96,PV97} mean field theories. 


\section{Outline of the theory: HFB formalism in coordinate space}

\begin{figure}[t]
\centering
\epsfig{file=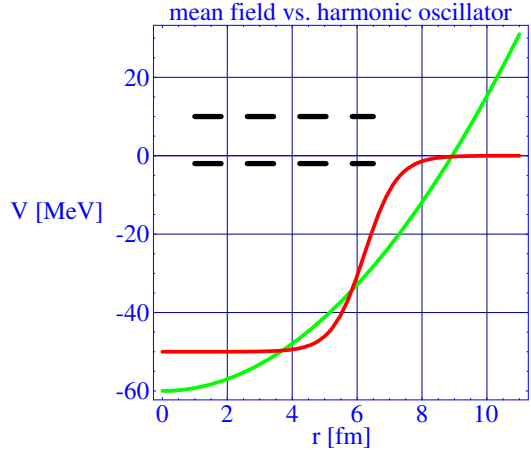,width=8.0cm} 
\caption{Inadequacy of shell model basis near the drip lines; need for
		 high-accuracy lattice representation.
         \label{fig:wsoscill}}
\end{figure}

As we move away from the valley of stability, surprisingly little is
known about the pairing force: For example, what is its density
dependence? Large pairing correlations are expected near the drip
lines which are no longer a small residual interaction. Neutron-rich
nuclei are expected to be highly superfluid due to continuum
excitation of neutron ``Cooper pairs''. The Hartree-Fock-Bogoliubov
(HFB) theory unifies the HF mean field theory and the BCS pairing
theory into a single selfconsistent variational theory.
The main challenge in the theory of exotic nuclei
near the proton or neutron drip line is that the outermost nucleons
are weakly bound (which implies a very large spatial extent), and that
the weakly-bound states are strongly coupled to the particle
continuum. This represents a major problem for mean field theories
that are based on the traditional shell model basis expansion method
in which one utilizes bound harmonic oscillator basis wavefunctions.
As illustrated in Figure \ref{fig:wsoscill} a weakly bound state can
still be reasonably well represented in the oscillator basis, but this
is no longer true for the continuum states. In fact, Nazarewicz et al.
\cite{NW94} have shown that near the driplines the harmonic oscillator
basis expansion does not converge even if $N=50$ oscillator quanta are
used. This implies that one either has to use a continuum-shell model
basis or one has to solve the problem directly on a coordinate space
lattice. We have chosen the latter method.

Several years ago, Umar {\it et al.} \cite{CU94} have developed a
three-dimensional HF code in Cartesian coordinates using the
Basis-Spline discretization technique. The program is
based on a density dependent effective N-N interaction (Skyrme force)
which also includes the spin-orbit interaction. The code has proven
efficient and extremely accurate; it incorporates BCS and
Lipkin-Nogami pairing, and various constraints. The configuration
space Hartree-Fock approach has had great successes in predicting
systematic trends in the global properties of nuclei, in particular
the mass, radii, and deformations across large regions of the periodic
table.

So far, our attempts to generalize this 3D code to include
self-consistent pairing forces (Hartree-Fock-Bogoliubov theory on the
lattice) have proven too ambitious. The reason may be the lack of a
suitable damping operator in 3D. We have therefore taken a different
approach and developed a new Hartree-Fock + BCS pairing code in
cylindrical coordinates for axially symmetric nuclei, based on the
Galerkin method with B-Spline test functions \cite{KO96,K96}. The
new code has been written in Fortran 90 and makes extensive use of new
data concepts, dynamic memory allocation and pointer variables.
Extending this code, we believe that it will be easier to implement
HFB in 2D because one can use highly efficient LAPACK routines to
diagonalize the lattice Hamiltonian and does not necessarily rely on a
damping operator.

We outline now our basic theoretical approach for lattice HFB.
As is customary, we start by expanding the nucleon field operator
into a complete orthonormal set of s.p. basis states $\phi_i$
\begin{equation}
\hat{\psi}^\dagger ({\bf r},s) = \sum_i \ \hat{c}_i^\dagger \ \phi_i^* ({\bf r},s)
\end{equation}
\noindent which leads to the Hamiltonian in occupation number representation
\begin{equation}
\hat{H}= \sum_{i,j} < i|\ t\ |j> \ \hat{c}_i^\dagger \ \hat{c}_j \ +
\frac{1}{4} \sum_{i,j,m,n} <ij|\ \tilde{v}^{(2)} \ |mn>
\ \hat{c}_i^\dagger \ \hat{c}_j^\dagger \ \hat{c}_n \ \hat{c}_m	\ .
\end{equation}
\noindent Like in the BCS theory, one performs a canonical transformation
to quasiparticle operators $\hat{\beta},\hat{\beta}^\dagger$
\begin{equation}
\left( 
\begin{array}{c}
\hat{\beta} \\ 
\hat{\beta}^\dagger 
\end{array}
\right) =
\left( 
\begin{array}{cc}
 U^\dagger & V^\dagger \\ 
  V^T & U^T 
\end{array}
\right) 
\left( 
\begin{array}{c}
\hat{c} \\ 
\hat{c}^\dagger
\end{array}
\right) \ .
\end{equation}
\noindent The HFB ground state is defined as the quasiparticle vacuum
\begin{equation}
\hat{\beta}_k \ | \Phi_0 > \ = \ 0 \ .
\end{equation}
\noindent The basic building blocks of the theory are the normal density
\begin{equation}
\rho_{ij} = < \Phi_0 | \hat{c}_j^\dagger \ \hat{c}_i | \Phi_0 > = (V^*V^T)_{ij}
\end{equation}
\noindent and the pairing tensor
\begin{equation}
\kappa_{ij} = < \Phi_0 | \hat{c}_j \ \hat{c}_i | \Phi_0 > = (V^*U^T)_{ij}
\end{equation}
\noindent from which one can form the generalized density matrix
\begin{equation}
\mathcal{R} =
\left( 
\begin{array}{cc}
 \rho & \kappa \\ 
 - \kappa^* & 1 - \rho^* 
\end{array}
\right)
\ \Rightarrow \mathcal{R}^\dagger = \mathcal{R}, \ \mathcal{R}^2 = \mathcal{R} \ .
\end{equation}
\noindent Using the definition of the HFB ground state energy
\begin{equation}
E' ( \mathcal{R} ) = < \Phi_0 | \hat{H} - \lambda \hat{N} | \Phi_0 >
\end{equation}
\noindent we derive the equations of motion from the variational principle
\begin{equation}
\delta \ [ E'( \mathcal{R} ) - {\rm{tr}} \ \Lambda ( \mathcal{R}^2 - \mathcal{R} ) ] \ = \ 0
\end{equation}
\noindent which results in the standard HFB equations
\begin{equation}
[ \mathcal{H}, {\mathcal{R}} ] \ = \ 0 
\end{equation}
\noindent with the generalized single-particle Hamiltonian
\begin{equation}
\mathcal{H} =
\left( 
\begin{array}{cc}
 h & \Delta \\ 
 - \Delta^* & -h^* 
\end{array}
\right)
; \  h = \partial E' / \partial \rho, \ \Delta = \partial E' / \partial \kappa^* \ .
\end{equation}

Our goal is to transform to a coordinate space representation and solve the
resulting differential equations on a lattice. For this purpose, we define two types of 
quasi-particle wavefunctions $\phi_1,\phi_2$
\begin{equation}
\phi_1^* (E_n, {\bf r} \sigma) \ = \ \sum_i U_{in} \ \phi_i ({\bf r} \sigma) \ ,\ \ \ 
\phi_2 (E_n, {\bf r} \sigma) \ = \ \sum_i V_{in}^* \ \phi_i ({\bf r} \sigma)
\end{equation}
\noindent in terms of which the particle density matrix for the HFB ground state assumes a very
simple mathematical structure \cite{DN96}
\begin{eqnarray}
\rho_0 \ ({\bf r}, \sigma, {\bf r}', \sigma' ) \ = \ 
< \Phi_0 | \ \hat{\psi}^\dagger ({\bf r}' \sigma') \ \hat{\psi} ({\bf r} \sigma) \ | \Phi_0 > \nonumber \\
= \sum_{i,j} \rho_{ij} \ \phi_i ({\bf r} \sigma) \ \phi_j^* ({\bf r}' \sigma')
= \sum_{E_n > 0}^{\infty} 
\phi_2 (E_n, {\bf r} \sigma) \ \phi_2^* (E_n, {\bf r}' \sigma') \ .
\end{eqnarray}
\noindent In a similar fashion we obtain for the pairing tensor
\begin{eqnarray}
\kappa_0 \ ({\bf r}, \sigma, {\bf r}', \sigma' ) \ = \ 
< \Phi_0 | \ \hat{\psi} ({\bf r}' \sigma') \ \hat{\psi} ({\bf r} \sigma) \ | \Phi_0 > \nonumber \\
= \sum_{i,j} \kappa_{ij} \ \phi_i ({\bf r} \sigma) \ \phi_j ({\bf r}' \sigma')
= \sum_{E_n > 0}^{\infty} \phi_2 (E_n, {\bf r} \sigma) \ \phi_1^* (E_n, {\bf r}' \sigma') \ .
\end{eqnarray}

For certain types of effective interactions (e.g. Skyrme forces) the
HFB equations in coordinate space are local and have a structure
which is reminiscent of the Dirac equation \cite{DN96}
\begin{equation}
\left( \matrix{ ( h-\lambda ) & \tilde h \cr 
                  \tilde h & - ( h-\lambda ) \cr} \right)
\left( \matrix{ \phi_1({\bf r}) \cr
                \phi_2({\bf r}) \cr} \right) = E
\left( \matrix{ \phi_1({\bf r}) \cr
                \phi_2({\bf r}) \cr} \right) \ ,
\label{eq:hfbeqn}
\end{equation}
where $h$ is the ``particle'' Hamiltonian and $\tilde h$ denotes the
``pairing'' Hamiltonian.

The various terms in $h$ depend on the particle densities $\rho_q(\bf r)$
for protons and neutrons, on the kinetic energy density $\tau_q(\bf r)$,
and on the spin-current tensor $J_{ij}(\bf r) $. The
pairing Hamiltonian $\tilde h$ has a similar structure, but depends on
the pairing densities $\tilde \rho_q(\bf r), \tilde \tau_q(\bf r)$
and $\tilde J_{ij}(\bf r)$ instead. Because of the
structural similarity between the Dirac equation and the HFB equation in
coordinate space, we encounter here similar computational challenges: for
example, the spectrum of quasiparticle energies $E$ is unbounded from above {\em
and} below. The spectrum is discrete for $|E|<-\lambda$
and continuous for $|E|>-\lambda$. In the case of axially symmetric nuclei, the
spinor wavefunctions $\phi_1({\bf r})$ and $\phi_2({\bf r})$ have the structure
\begin{equation}
\psi^\Omega (\phi,r,z) = \frac{1}{\sqrt{2 \pi}}
\left( \matrix{ e^{i(\Omega - \frac 12)\phi} \ U(r,z) \cr 
                e^{i(\Omega + \frac 12)\phi} \ L(r,z) \cr} \right) \ .
\label{eq:spinor}
\end{equation}


\section{Computational method: Spline-Galerkin lattice representation}

\begin{figure}[t]
\centering
\epsfig{file=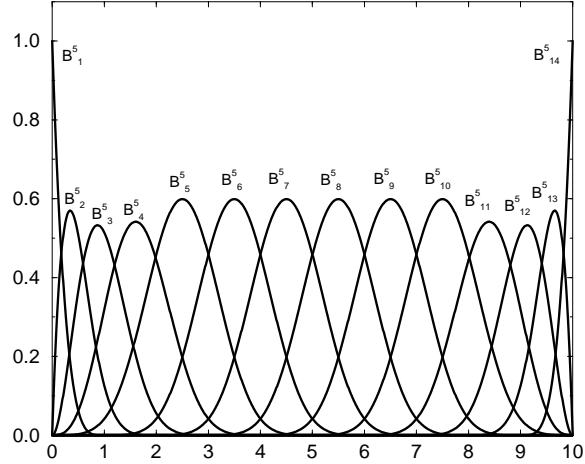,width=8.0cm} 
\caption{Set of fifth-order B-Splines for fixed boundary conditions.
         \label{fig:bspline}}
\end{figure}

For nuclei near the p/n driplines, we overcome the convergence
problems of the traditional shell-model expansion method by
representing the nuclear Hamiltonian on a lattice utilizing a
Basis-Spline expansion \cite{WO95,K96,KO96}. B-Splines $B_i^M(x)$ are
piecewise-continuous polynomial functions of order $(M-1)$. They
represent generalizations of finite elements which are B-splines with
$M=2$. A set of fifth-order B-Splines is shown in Figure \ref{fig:bspline}.

Let us now discuss the Galerkin method with B-Spline test
functions. We consider an arbitrary (differential) operator equation
\begin{equation}
{\cal{O}} \bar{f}(x) - \bar{g}(x) = 0\;.
\label{eq:eq4}
\end{equation}
Special cases include eigenvalue equations of the HF/HFB type where ${\cal{O}}=h$
and $\bar{g}(x)=E \bar{f}(x)$.
We assume that both $\bar{f}(x)$ and $\bar{g}(x)$ are well approximated by Spline
functions
\begin{equation}
\bar{f}(x) \approx f(x) \equiv \sum_{i=1}^{\cal N} B_i^M(x)a^i\;,\ \ \ \ 
\bar{g}(x) \approx g(x) \equiv \sum_{i=1}^{\cal N} B_i^M(x)b^i\;.
\label{eq:eq5}
\end{equation}
Because the functions $f(x)$ and $g(x)$ are approximations to the exact functions
$\bar{f}(x)$ and $\bar{g}(x)$, the operator equation will in general only be
approximately fulfilled
\begin{equation}
{\cal{O}} f(x) - g(x) = R(x)\;.
\label{eq:eq6}
\end{equation}
The quantity $R(x)$ is called the {\it residual}; it is a measure of the accuracy
of the lattice representation. We multiply the last equation from the left with
the spline function $B_k(x)$ and integrate over $x$
\begin{equation}
\int v(x) dx B_k(x) {\cal{O}} f(x) - \int v(x) dx B_k(x) g(x) = 
\int v(x) dx B_k(x) R(x)\;\;.
\label{eq:eq7}
\end{equation}
We have included a volume element weight function $v(x)$ in the integrals
to emphasize 
that the formalism applies to arbitrary curvilinear coordinates.
Various schemes exist to minimize the residual function $R(x)$; in the Galerkin
method one requires that there be no overlap between the
residual and an arbitrary B-spline function
\begin{equation}
\int v(x) dx B_k(x) R(x) = 0 \;\;.
\label{eq:eq8}
\end{equation}
This so called {\it Galerkin condition} amounts to a {\it global
reduction of the residual}. Applying the Galerkin condition and
inserting the B-Spline expansions for $f(x)$ and $g(x)$ results in
\begin{equation}
\sum_i \left[\int v(x) dx B_k(x) {\cal{O}} B_i(x) \right] a^i - 
\sum_i \left[\int v(x) dx B_k(x) B_i(x) \right] b^i = 0\;\;.
\label{eq:eq9}
\end{equation}
Defining the matrix elements
\begin{equation}
{\cal{O}}_{k i}= \int v(x) dx B_k(x) {\cal{O}} B_i(x)\;\;,\ \  
G_{k i} = \int v(x) dx B_k(x) B_i(x)\;
\label{eq:eq10}
\end{equation}
transforms the (differential) operator equation into a matrix $\times$ vector
equation 
\begin{equation}
\sum_i {\cal{O}}_{k i} a^i = \sum_i G_{k i} b^i\;
\label{eq:eq11}
\end{equation}
which can be implemented on modern vector or parallel computers with
high efficiency. The matrix $G_{k i}$ is sometimes referred to as the
{\it Gram} matrix; it represents the nonvanishing overlap integrals
between different B-Spline functions (see Fig. \ref{fig:bspline}). We
eliminate the expansion coefficients $a^i,\ b^i$ in the last equation
by introducing the function values at the lattice support points
$x_\alpha$ including both interior and boundary points.

The upper $(U)$ and lower $(L)$ components of the spinor wavefunctions
defined earlier are represented on the 2-D lattice $(r_\alpha,
z_\beta)$ by a product of Basis Splines $B_i (x)$ evaluated at the
lattice support points
\begin{equation}
U(r_\alpha, z_\beta) = \sum_{i,j} B_i (r_\alpha) \ B_j (z_\beta) \ U^{ij} \ ,
\ \ \ 
L(r_\alpha, z_\beta) = \sum_{i,j} B_i (r_\alpha) \ B_j (z_\beta) \ L^{ij}
\ .
\label{eq:eq16}
\end{equation}
We are also extending our previous B-spline work to include nonlinear grids. Use
of a nonlinear lattice should be most useful for loosely bound systems near the
proton or neutron drip lines. Non-Cartesian coordinates necessitate the use of
fixed endpoint boundary conditions; much effort has been directed toward
improving the treatment of these boundaries \cite{KO96}.


\section{Numerical tests and results}
\begin{figure}[t]
\centering
\epsfig{file=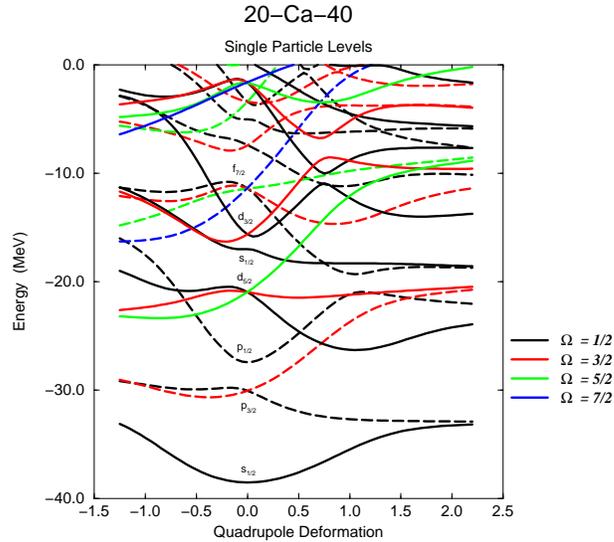,width=8.0cm} 
\caption{Single-particle neutron levels for $^{40}_{20}Ca$ as function of
         quadrupole deformation in the Woods-Saxon shell model. Solid lines
		 indicate positive parity and broken lines negative parity levels.
         \label{fig:ca40}}
\end{figure}

We expect our Spline techniques to be superior to the traditional
harmonic oscillator basis expansion method in cases of very strong
nuclear deformation. To illustrate this point, we have performed a
numerical test using a phenomenological (Woods-Saxon) deformed shell
model potential. We calculate the single-particle energy spectrum for
neutrons in $^{40}Ca$ for quadrupole deformations ranging from strong
oblate ($\beta_2=-1.25$) to extreme prolate ($\beta_2=+2.25$). The
results are shown in Fig. \ref{fig:ca40}. Apparently,
for $\beta_2=0$ we correctly reproduce the spherical shell structure
of magic nuclei. As $\beta_2$ approaches large positive values our
s.p. potential approaches the structure of two separated potential
wells; as expected, we observe pairs of levels with opposite parity
that are becoming degenerate in energy. The largest quadrupole
deformation corresponds physically to a symmetric fission
configuration. Clearly, such configurations cannot be described in a
single oscillator basis, which confirms the numerical superiority of
the B-Spline lattice technique.

\begin{figure}[t]
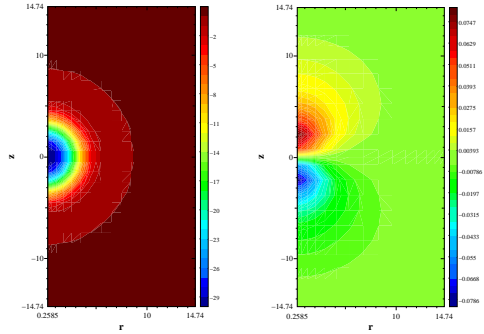

\centering
\epsfig{file=lipot2.epsi,width=1.26in,angle=0}
\epsfig{file=lin3u.epsi,width=1.33in,angle=0}
\caption{ (a) Woods Saxon potential for $^{11}_{\ 3}$Li; (b) spin-up 
component of the wave function for the last bound neutron.
         \label{fig:li11}}
\end{figure}

In a second test calculation, we have investigated the properties of a nucleus
near the neutron drip line. During the last decade the discovery of a `neutron
halo' in several neutron-rich isotopes generated a great deal of interest in the
area of weakly bound quantum systems. The halo effect was first observed in
$^{11}_{\ 3}$Li, which consists of three protons and six neutrons in a central
core and two planetary neutrons which comprise the halo. By adjusting the depth
of the Woods-Saxon potential so that the separation energy of the last bound
neutron is only $10$ keV, i.e. very close to the continuum, we were able to
determine this neutron wavefunction on the lattice which shows a very large
spatial extent (see Fig. \ref{fig:li11}). We conclude that the B-Spline lattice techniques are
well-suited for representing weakly bound states near the drip lines; a similar
calculation in the basis expansion method would require a large number of
oscillator shells.

\begin{figure}[t]
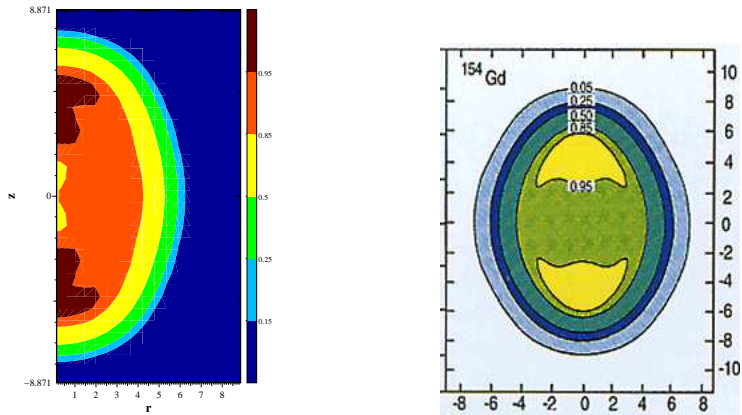

\centering
\epsfig{file=gd154rhop.epsi,width=4.0cm}
\ \ \ \ \ \ \ \ \ \ \ \ \ \ \ \ \ \ 
\epsfig{file=gd154.epsi,width=4.0cm}
\caption{Proton density for $^{154}_{\ 64}$Gd, (a) calculated with the 2-D
Skyrme Hartree-Fock + BCS pairing code; (b) measured charge distribution.
         \label{fig:gd154}}
\end{figure}

We now discuss our numerical results for the selfconsistent
Hartree-Fock calculations with Skyrme-M$^*$ interaction and BCS
pairing. This is a special case of the HFB equation with a constant
pairing matrix element. In Fig. \ref{fig:gd154} we display the proton
density for a heavy nucleus, $^{154}_{\ 64}$Gd, calculated with our
new 2-D (HF+BCS) code. It should be noted that ALL 154 nucleons are
treated dynamically (no inert core approximation). The theoretical
charge density looks quite similar to the experimental result which is
shown on the right hand side.

For several spherical nuclei, we have also compared the selfconsistent
s.p. energy levels of our 2-D Spline-Galerkin code with a fully
converged 1-D radial calculation. The result is shown in Table 1.

\begin{table}[t]
\caption{Total binding energy and s.p. energy levels for $^{16}O$,
calculated in the Hartree-Fock + BCS pairing theory with the SkM$^*$ force.
Comparison of results from 1D radial finite difference code 
\protect\cite{PGF77} and our
new 2D Spline-Galerkin code. We obtain the same level of
accuracy despite the $25$ times larger lattice spacing $\Delta=0.625$ fm.}
\begin{center}
\footnotesize
\begin{tabular}{lcrcrc}
\ \\
\hline \\
\  & 1D Radial  & 2D Spline-Galerkin \\
\  & $\Delta=0.025$fm & $\Delta=0.625$fm \\
\hline \\
$E_{tot}$      & -127.73 MeV  & -127.48 MeV  \\
$E_{s1/2}(n)$  &  -33.31 MeV  &  -33.29 MeV  \\
$E_{p3/2}(n)$  &  -19.88 MeV  &  -19.86 MeV  \\
$E_{p1/2}(n)$  &  -13.55 MeV  &  -13.53 MeV  \\
$E_{s1/2}(p)$  &  -29.74 MeV  &  -29.72 MeV  \\
$E_{p3/2}(p)$  &  -16.48 MeV  &  -16.45 MeV  \\
$E_{p1/2}(p)$  &  -10.27 MeV  &  -10.26 MeV  \\
\hline
\end{tabular}
\end{center}
\end{table}


\subsection{Plans and Future Directions}
Having validated our new (HF+BCS) code on a 2D lattice with the
Spline-Galerkin method, we plan to proceed as follows: We are
currently working on the 2D HFB implementation with a pairing
delta-force. After that, we will generalize the code utilizing the
full SkP force with mean pairing field and pairing spin-orbit term. We
will also add appropriate constraints, e.g. $Q_{20},Q_{30},\omega j_x$
for calculating potential energy surfaces and rotational bands. As we
compare the observables (e.g. total binding energy, charge radii,
densities $\rho_{p,n}({\bf r})$, separation energies for neutrons and
protons, pairing gaps) with experimental data from the RIB facilities,
it will almost certainly be necessary to develop new effective N-N
interactions as we move farther away from the stability line towards the
p/n drip lines.


\section*{Acknowledgments}
This research project was sponsored by the U.S. Department of Energy
under contract No. DE-FG02-96ER40975 with Vanderbilt University. Some of the
numerical calculations were carried out on CRAY supercomputers at the National
Energy Research Scientific Computing Center (NERSC) at Lawrence Berkeley
National Laboratory. We also acknowledge travel support from the NATO
Collaborative Research Grants Program.



\end{document}